\documentstyle[aps,prl,multicol,epsf,latexsym]{revtex}

\begin{document}
\draft

\title{Nonequilibrium Phase Transitions in 
Directed Small-World Networks}

\author{Alejandro D. S\'anchez\cite{mail1}, 
Juan M. L\'opez\cite{mail2} 
and Miguel A. Rodr\'{\i}guez\cite{mail3}}
\address{Instituto de F\'{\i}sica de Cantabria, 
CSIC-UC, E-39005 Santander, Spain}

\maketitle

\begin{abstract}
Many social, biological, and economic systems can be 
approached by complex networks of interacting  
units. The behaviour of several 
models on small-world networks 
has recently been studied. These models are expected to
capture the essential features of the complex 
processes taking place on real networks like       
disease spreading, formation of public opinion,
distribution of wealth, etc. In many
of these systems relations are {\em directed}, in the sense
that links only act in one direction (outwards or inwards).
We investigate the effect of directed links on the behaviour 
of a simple spin-like model evolving on a small-world network. 
We show that directed networks may lead to a highly nontrivial 
phase diagram including first and second-order phase transitions  
out of equilibrium. 
\end{abstract}

 %Revised

\begin{multicols}{2}
\narrowtext
Complex networks have recently attracted an increasing 
interest among physicists. The main reason being that 
they seem to be exceedingly simple model systems of complex
behaviour in real world networks \cite{watts-book,amaral}, 
including chemical reaction
networks \cite{alon}, food webs \cite{pimm,paine,cann}, 
the Internet \cite{huberman,baraba} and the 
World-Wide-Web \cite{guido}, 
metabolic \cite{jeong1} and
protein networks\cite{jeong2}, 
scientific collaboration networks \cite{newman1}, etc. The hope
is that the ideas and techniques developed in the last
fifty years in the field of statistical physics to deal with
cooperative phenomena in many body systems,
may be useful to understand emergent complex behaviour
in systems outside the traditional realm of physics. 
In particular, small-world (SW) networks, recently
introduced by Watts and Strogatz \cite{watts}, have been very
much studied because they constitute an interesting attempt
to translate the complex topology of social, economic and
physical networks into a simple model. SW networks result from
randomly replacing a fraction $p$ of links of a $d$-dimensional regular
lattice with new random links. As a result of this random rewiring,
SW networks interpolate between the two limiting cases of a regular 
lattice ($p=0$) and completely random graphs ($p=1$).
Studies of real network data have shown
that SW-like topologies are found in situations
as diverse as the network of movie actors
collaboration, the electric power grid of Southern California,
the network of world airports, the acquaintance network
of Mormons, etc \cite{watts-book,amaral}.

Many topological properties of the SW model
have recently been investigated, as for instance 
the shortest-path distance and clustering 
coefficient \cite{watts,barrat},
the crossover from regular to SW behaviour  
occurring at $p=0$ \cite{barthelemy},
a mean-field solution \cite{newman}, and percolation on 
SW networks \cite{moore}
among others. Specifically, SW models are expected to play
an important role in understanding the interplay between
the underlying disordered
network and the dynamics of many social or economic processes,
such as distribution of wealth, disease 
spreading, transmission of cultural traits, 
and formation of public opinion \cite{watts-book}. 

In the language of social network analysis \cite{social-book}, 
sites are referred to as {\em actors}. Actors may represent
individuals, companies, airports, countries, etc., depending
on the social or economic process we are interested in.
Actors are linked to one another by a relational, social
or physical {\em tie} like
for instance, friendship, business transactions,
flight connections, kinship, or
scientific collaboration, among many others. 
Some of those relational links are {\em symmetric}, 
in the sense that if Alice is tied to Bob, then Bob must 
also be tied to Alice, as occurs for instance in 
the authorship of scientific papers. 
However, many other networks are {\em directed} 
and exhibit links that are definitely {\em asymmetric}, 
like for instance in the case of networks of
import-export of goods, world-wide-web pages links, 
lending transactions, food webs, cultural influences, etc.  
In directed networks then,
when Alice is tied to Bob, Bob may not be linked to 
Alice but to someone else instead. Asymmetric synaptic 
strengths have already been shown to be very important 
in trying to describe the process of learning in realistic 
neural network model approaches to brain 
function\cite{parisi,crisanti}.

Several models have recently been studied
in order to understand the effect of SW topology 
on classical systems like the Ising 
model \cite{barrat} or the
spread of infections and epidemics \cite{chule,romu}. 
Such a simple models are expected to capture the essential 
features of the more complicated processes taking place on 
real networks. However, as mentioned earlier, many social,  
commercial or biological relations are asymmetric and the following 
question naturally arises: What is the effect of directed 
links on a simple model that evolves on the network?
 
In this Letter we investigate the effect of directed SW topology 
on the behaviour of a simple model. In case 
only undirected links are used, our model becomes identical to 
the classical Ising model on a standard (undirected) SW 
network \cite{barrat}.
This allows us to study, in a systematic way,
the effect of directed ties on this classical model.
We find that the existence of directed links 
completely changes the behaviour of the system from
mean-field behaviour (for undirected networks) to
a highly nontrivial and rich phase diagram in the case of
directed networks.
By means of extensive numerical simulations 
we find that, for rewiring probabilities in 
the range $0 < p < p_c$, 
the model exhibits a line of continuous phase 
transitions from an ordered to a disordered state. Those phase
transitions occur at a critical value of the temperature $T_c(p)$,
which depends on $p$. However, for higher disorder 
densities $p_c < p \leq 1$, the phase transition becomes
first-order.
Our results show that, in order to model biological, social
or economic processes on complex networks,
it is crucial to take into account the character, 
directed or undirected, of the corresponding relational links. 

{\em The model.-}  
We have studied directed networks in $d=1$ and 2.
For simplicity we focus here on $d=2$ and further results 
in $d=1$ will be published elsewhere. 
In order to construct a directed SW network, 
we start from a 2-dimensional square lattice consisting
of sites linked to their four nearest-neighbours by
both, outgoing and incoming links. 
Then, with probability $p$ we reconnect nearest-neighbour
outgoing links to a different site chosen at random. 
After repeating this process for every outgoing link we are left
with a network with a density $p$ of SW directed links,
as shown in Figure 1. 
Note that, by this procedure, every site will have exactly
four outgoing links and a varying (random) number of
incoming links. Generalization to higher dimensions is  
straightforward.

Adopting social network nomenclature,
actors are then placed at the network sites. 
Any given actor is connected by four outgoing links to
other actors, which we call {\em mates}. 
We allow every actor to be in one of two possible
states, so that, at any given time, the state of an
actor is described by a binary spin-like variable
$s_i \in \{+1,-1\}$. Depending on the state of their
mates, an actor may change its state
according to a majority (ferromagnetic) rule: 
Actors prefer to be in the same state as their mates are. 
In order to implement this, we introduce the pay-off function 
\begin{equation}
G(i)=2 s_i\sum _{{\rm mates \: of}\: i} s_j,
\label{pay-off}
\end{equation}
where the sum is carried out over the four mates of 
actor $i$. Note that this pay-off function is positive
whenever $s_i$ points in the same direction as the majority of
its four mates.
External noise is included to allow  
some degree of randomness in the time evolution by means of
a temperature-like parameter, $T$, which we shall call 
temperature for short from now on.
For a given value of the external temperature, 
the update of the model is then performed as follows:
At each time step, an actor (network site) 
is randomly chosen and 
its corresponding pay-off function $G(i)$ is computed according
to Eq. (\ref{pay-off}). If $G(i) < 0$, actor $i$ is opposing 
its mates' majority and the change $s_i \to -s_i$ is accepted.
Unfavorable changes, {\it i.e.} when $G(i) > 0$, are accepted with 
probability $\exp[-G(i)/T]$, which
depends on temperature in the usual fashion. 

Concerning the physics of the above defined model, 
there are two interesting points that should be  
explicitly mentioned. On the one hand,  
the model is nonequilibrium since detailed balance  
is not satisfied. On the other hand, 
the model is not simply the asymmetric  
counterpart of the Ising model, since the pay-off  
function $G(i)$ in Eq. (\ref{pay-off}) does not include  
the corresponding interaction 
terms coming from the ingoing links (needed 
in order to identify $G(i)$ with the energy change
after a spin update in the asymmetric Ising model). In fact, 
one can easily see that the pay-off function $G(i)$  
of our model cannot be written as variation of  
any Hamiltonian. 
However, we would like to remark that if only symmetric 
links are allowed, our model becomes
exactly equal to the (equilibrium) Ising model in 
an undirected SW network that was studied 
in Ref.\cite{barrat}.
This can be seen by a simple comparison of
the pay-off function Eq. (\ref{pay-off}) with the
change of energy after a spin update in the
standard (symmetric) Ising 
ferromagnet. In this case, it is known that
the system presents mean-field behaviour for
any value of the disorder $p>0$ \cite{barrat}.

{\em Results.-}
We have carried out extensive numerical simulations 
of the model for different values of the 
density of SW directed ties $p$ and temperature.
Our results are qualitatively the same for
directed networks generated from regular lattices
in $d=1$ and $2$. In the following we focus on
$d=2$.
We have simulated the model in directed SW networks 
generated from $L \times L$ square lattices for sizes
ranging from $L=8$ to $100$ and different 
rewiring probabilities $p \in [0,1]$. 
The system is left to evolve until, after some transient, 
a stationary nonequilibrium state is reached. 
The stationary state can be described by the 
appropriate 
order parameter, which can be defined in a 
natural way
by means of the 'magnetization' per site
\begin{equation}
\label{m}
m = {1 \over L^2} \sum_{i=1}^{L^2} s_i. 
\end{equation}
We find that the system becomes ordered, 
{\it i.e.} $\langle |m| \rangle \neq 0$, 
below a critical temperature $T_c(p)$, so that
most actors are, in average, in the same state.  
In Figure 2 the average absolute value of the order 
parameter is plotted {\it vs.} temperature for two 
different values of the disorder $p=0.1$ and 0.9,
calculated in systems of different sizes. 
For every system size $L^2$,
results were averaged over both, $10$ runs
of the dynamics for each network
realization and $n$ different realizations of
the network, in such a way that 
$n \times L^2 \approx 1.5 \times 10^5$.
Figure 2 shows that the order-disorder transition 
is continuous (top panel) for a low disorder 
density, while it becomes discontinuous for a 
higher concentration of directed SW links (bottom panel). 
A more systematic study of the phase diagram, 
as shown in Figure 3, reveals that there is a line of 
continuous phase transitions for disorder 
densities below some critical value $p_c$.
Very interestingly, the transition becomes first-order
above $p_c$, indicating that there exists a 
{\em nonequilibrium tricritical
point} at $p_c$. We estimate $p_c$ to be roughly 
at $p_c=0.65(5)$. The character, continuous or discontinuous, 
of the phase transition is better realized when looking at 
the probability density function (PDF) of the order parameter.
For sake of illustration, 
the insets of Fig. 3 show typical PDFs at points of 
the phase diagram (p,T), all near the critical line. 
From those PDFs, one can see that 
the phase transition is second-order from Fig. 3a to 3b, in the
region $p<p_c$. In contrast, for $p>p_c$ 
the transition is discontinuous, from Fig. 3c to 3d. 
The most probable values of $m$, which correspond to the 
two equally highest symmetric peaks in Fig. 3c, become unstable 
in favor of $m=0$ as the transition line is crossed 
towards Fig. 3d. The transition occurs in such a way 
that the order parameter exhibits a finite jump at the critical line. 

The critical behaviour of the model
in the region $p<p_c$, where transitions are continuous,
can be studied in detail. We find that the order parameter
exhibits finite-size scaling with exponents that depend on
the disorder density $p$. Close to the critical point, 
$|t| \to 0$,
we have $\langle |m| \rangle \sim |t|^\beta$, where 
$t = (T - T_c)/T_c$ is the reduced temperature. 
At the critical point, $t=0$,
the order parameter scales with system size as 
$\langle |m| \rangle \sim L^{-\beta/\nu}$, where $\nu$
is the correlation length exponent. Figure 4 displays
the behaviour of $\langle |m| \rangle$ {\it vs.} $L$ for
two disorder densities $p=0.1$ and 0.5 below $p_c \approx 0.65$.
Only for $T=T_c(p)$ a power-law is obtained and the slope 
of the straight line in a log-log plot gives an estimation 
of the ratio $\beta/\nu$ between critical exponents. From
Fig. 4 we obtain that for $p=0.1$ and $p=0.5$, 
$\beta/\nu=0.53(2)$ and $0.40(3)$, respectively. 
Moreover, from data collapse analysis (not shown) at the 
corresponding $T_c(p)$ we have $\beta=0.50(3)$, 
$\nu=0.94(6)$, and $\beta=0.30(3)$, 
$\nu=0.80(3)$ for $p=0.1$ and $p=0.5$, respectively. 
These critical exponents
are different from both mean-field, ($\beta = \nu = 1/2$) 
and exact values ($\beta = 1/8$ and $\nu = 1$) for the
Ising model in $d=2$.

{\em Conclusions.-} 
Many social, economic and biological networks
in the real world exhibit directed ties or
relations. This may be modeled by including 
directed links in the corresponding complex 
network model. In addition, spin-like models, 
borrowed from statistical physics, have recently 
been proposed as toy models to understand 
some social processes, like for instance
conflict vs. cooperation
among coalitions \cite{galam,galam2} 
or formation of cultural domains \cite{claudio}.
We claim that the directedness
of the network may strongly affect the behaviour 
of simple processes evolving on complex networks. 
We studied the effect of a directed small-world 
topology on a very simple spin model. 
Our model becomes equal to the Ising model when 
all links are undirected. From numerical simulations 
we showed that, when directed links exist, 
the phase diagram of 
the model is nontrivial. We found that the system exhibits
continuous phase transitions for disorder densities
below a critical threshold $p_c \approx 0.65$. 
For stronger disorder, the transition is
first-order. At this stage we can only speculate 
that the competition among weekly coupled 
clusters may be related
to the existence of first-order transitions.    

We believe that the effect of directed 
links may be relevant in 
other types of disordered networks, like free-scale
networks, and different dynamical models.
In trying to model real systems, directed links
may play an important and unforeseen role.     
 
{\em Acknowledgement.-} 
We would like to thank J. G\"uemez for collaboration
at the early stages of this work. We are also grateful to
J.J. Ramasco, L. Pesquera and X. Guardiola
for stimulating discussions and comments.
A.D.S. acknowledges postdoctoral fellowship from 
CONICET (Argentina).
J.M.L. acknowledges financial support from the 
European Commission (contract HPMF-CT-1999-00133). 
Partial financial support from Ministerio de Ciencia y 
Tecnolog{\'\i}a (Spain) through projects 
BFM2000-0628-C03-02 and BFM2000-1105 is acknowledged.

\begin{figure}
\centerline{
\epsfxsize=5.0cm
\epsfbox{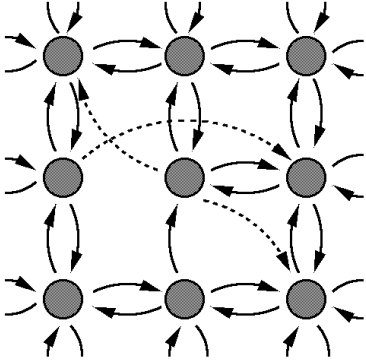}}
\caption{Sketch of a directed small-world network constructed from
a square regular lattice in $d=2$. For sake of clarity only a few links
have been reconnected. Arrows indicate the direction
of the corresponding link. Dotted lines represent rewired links.
Note that every site always has four outgoing links.}
\end{figure}

\begin{figure}
\centerline{
\epsfxsize=7.5cm
\epsfbox{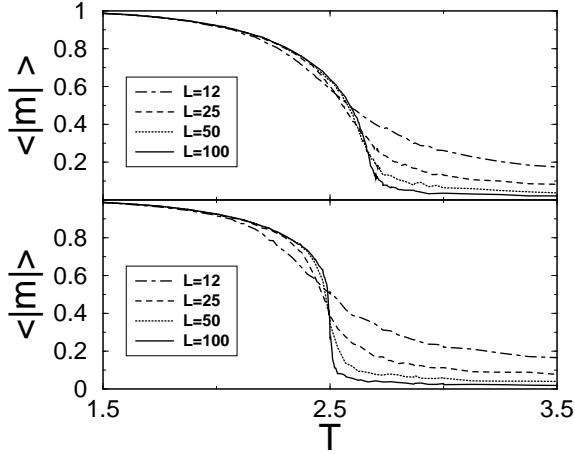}}
\caption{Order parameter {\it vs.} $T$ for different system
sizes. For $p=0.1$ (top panel) the transition is continuous.
For a higher disorder density, $p=0.9$, the transition becomes 
first-order (bottom panel).}
\end{figure}

\begin{figure} 
\centerline{
\epsfxsize=7.5cm
\epsfbox{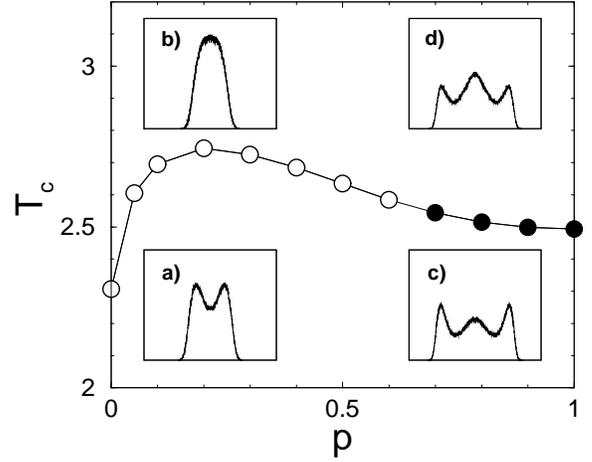}}
\caption{Phase diagram of the model. The system
is in the ordered state below the line. Points
are numerical determinations of the critical 
temperatures $T_c(p)$ for different degrees of disorder. 
The transition is continuous for small values of
$p$ (circles), while it becomes discontinuous 
for $p$ larger than $p_c \approx 0.65$ 
(filled circles). The insets show the  
PDFs of $m$ for: a) $p=0.1$, $T=2.68$;
b) $p=0.1$, $T=2.70$; c) $p=0.9$, $T=2.498$; d) $p=0.9$, 
$T=2.500$. Simulations were performed
in a $100 \times 100$ sites network.}
\end{figure}

\begin{figure}
\centerline{
\epsfxsize=7.5cm
\epsfbox{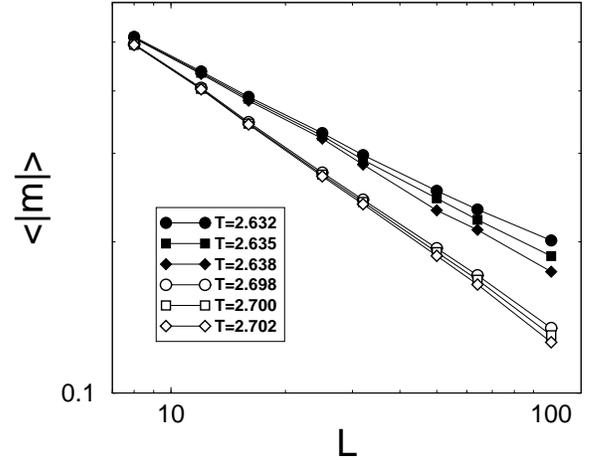}}
 \caption{Finite-size scaling of the order parameter for
 $p=0.1$ (hollow symbols) and $p=0.5$ (filled symbols),
 both below $p_c$. Power-law behaviour is obtained for
 $T=2.700(2)$ and $T=2.635(3)$ for $p=0.1$ (squares) 
 and $p=0.5$ (filled squares), respectively.}
\end{figure}

\end{multicols}
\end{document}